\documentclass{article}
\usepackage{hyperref}
\usepackage{datetime}
\usepackage{caption}
\usepackage{amsmath}

\newif\ifpdf
\ifx\pdfoutput\undefined
\pdffalse 
\else
\pdfoutput=1 
\pdftrue
\fi

\ifpdf
\usepackage[pdftex]{graphicx}
\else
\usepackage{graphicx}
\fi

\ifpdf
\DeclareGraphicsExtensions{.pdf, .jpg}
\else
\DeclareGraphicsExtensions{.eps, .jpg}
\fi

\begin{document}

\thispagestyle{empty}
{\vspace*{-2.8cm} \hspace*{8.5cm} \includegraphics[width=63mm]{lab_logo.pdf}}
\vspace{5.5cm}

{\noindent \large \textsf{Cheol E. Han, Luis R. Peraza, John-Paul Taylor, Marcus Kaiser}} \\
\\

{\noindent \LARGE \bf \textsf{Predicting age of human subjects based on\\ structural connectivity from diffusion tensor imaging}} \\
\vfill

{\noindent \LARGE \textsf{Technical Report No. 1}} \\
{\noindent \textsf{\today}} \\
{\hspace*{-3.17cm} \rule[3mm]{\textwidth}{0.75pt}} \\
{\LARGE \textsf{Dynamic Connectome Lab}} \\
\url{http://www.biological-networks.org/} 

\newpage
\thispagestyle{empty}

{ 
\noindent
\huge {\bf Predicting age of human subjects based on structural connectivity from diffusion tensor imaging} \\
\\
\normalsize

\vspace{1cm}

\noindent
Cheol E. Han$^{1,2}$,Luis R. Peraza$^3$,John-Paul Taylor$^3$, Marcus Kaiser$^{2,4,5}$ 


\vspace{1cm}

\noindent
\textit{$^1 $ Department of Biomedical Engineering, Korea University, South Korea \\ 	 $^2 $ Department of Brain Cognitive Sciences, Seoul National University, South Korea \\	 $^3 $ Institute for Ageing and Health, Newcastle University, UK \\	 $^4 $ School of Computing Science, Newcastle University, UK \\	 $^5 $ Institute of Neuroscience, Newcastle University, UK \\
\\
E-Mail: \href{mailto:cheolhan@gmail.com}{cheolhan@gmail.com}\\
\url{http://bia.korea.ac.kr/people/~cheolhan/} 
} \\

}

\section*{Summary}
Predicting brain maturity using noninvasive magnetic resonance images (MRI) can distinguish different age groups and help to assess neurodevelopmental disorders. However, group-wise differences are often less informative for assessing features of individuals. Here, we propose a simple method to predict the age of an individual subject solely based on structural connectivity data from diffusion tensor imaging (DTI). Our simple predictor computed a weighted sum of the strength of all connections of an individual. The weight consists of the fiber strength, given by the number of streamlines following tract tracing, multiplied by the importance of that connection for an observed feature---age in this case. We tested this approach using DTI data from 121 healthy subjects aged 4 to 85 years. After determining importance in a training dataset, our predicted ages in the test dataset showed a strong correlation ($\rho=0.77$) with real age deviating by, on average, only 10 years.


\newpage
\setcounter{page}{1}

%

%
%
\section{Introduction}
\label{sec:intro}

The study of how different components of the brain, may they be neurons or brain regions, are connected has become an emerging field within the neurosciences \cite{sporns, bullmore, kaiser2011}. Structural 
connectivity observes the physical wiring of neural circuits, while functional connectivity links brain regions with similar activity over time. Magnetic Resonance Imaging (MRI) facilitates human studies since 
it enables us to construct such networks noninvasively. At the macro scale where network nodes represent brain regions, Diffusion Tensor Imaging (DTI) allows us to quantify the number of streamlines after tract tracing as a proxy of connection strength between two regions.  
Analyzing brain networks can be a tool to understand the interaction between nodes and has already shown a strong correlation between cognitive functioning and global information integration for functional networks \cite{heuvel}.
Predicting brain maturity from MR images is beneficial to assess neurodevelopmental disorders. As pediatric disorders including ADHD have delays in brain maturity \cite{shaw2012}, investigating structural normality of pediatric brains can be a good screening strategy. Also in the elderly, it is beneficial to distinguish cognitive disorders from normal ageing. 
Taking advantage of the network analysis framework, classifying structural connectivity networks to distinguish different age ranges has previously been performed by looking at small sets of nodes, network motifs \cite{ribeiro2009}, or by sets of features for individual nodes, single-node motifs \cite{costa2009, echtermeyer}. Also network topological changes over age were investigated in structural connectivity \cite{gong, hagmann2010}. Whereas these studies use aggregate features of the network, we used only raw data about individual edges, as given by the number of streamlines following tract tracing of DTI data, to inform predictions of the age of a subject. While a link between functional connectivity and brain maturity was reported earlier \cite{dosenbach}, it remains unclear whether structural connections alone can be a comparatively suitable predictor as functional and structural connectivity are often related \cite{honey}. 
Here, we tested a simple prediction model for the age of a subject using information about structural connectivity based on diffusion tensor imaging. Our predictor computed a weighted sum of the structural connectivity matrix for each subject as a raw score, where the weight of each connection was predefined as the correlation coefficient between connection strength of that connection with age over all subjects in the training group. Then, we transformed the raw score to a predicted age with linear regression. We then evaluated our method over a test group, computing the correlation between predicted ages and the real ages in the test group.

\section{Methods}
\label{sec:methods}

\subsection{DTI database}

\begin{figure}[htb]

  \centering
  \centerline{\includegraphics[width=8.5cm]{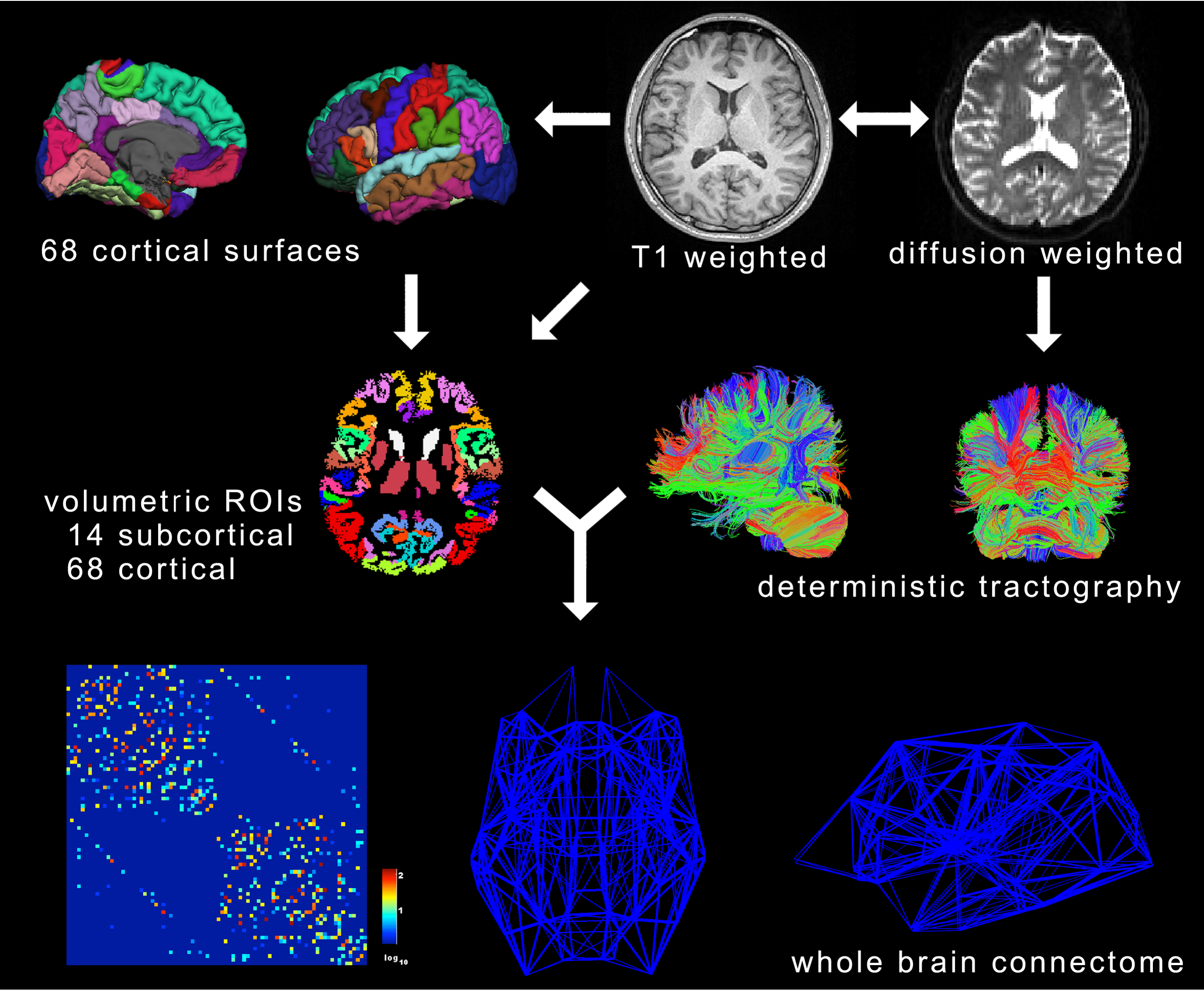}}
   \caption {DTI network inference.  }  \label{fig:networkflow}
\end{figure}

We made use of a public DTI-database (http://fcon\_1000.-projects.nitrc.org/indi/pro/nki.html) provided by the Nathan Kline Institute (NKI) \cite{nooner}. 
DTI-data were obtained with a 3 Tesla scanner (Siemens MAGNETOM TrioTim syngo, Erlangen, Germany). 
T1 weighted MRI data were obtained with 1mm isovoxel, FoV 256mm, TR=2500ms, and TE=3.5ms. 
DTI data were recorded with 2mm isovoxel, FoV=256mm, TR=10000ms, TE=91ms, and 64 diffusion directions with b-factor of 1000 s mm$^{-2}$ and 12 b$_0$ images. 
We included 121 participants between 4 and 85 years. We used Freesurfer to obtain surface meshes of the boundary between gray and white matter from T1 anatomical brain images (http://surfer.nmr.-mgh.harvard.edu), 
see Figure~\ref{fig:networkflow}. 
After registering surface meshes into the DTI space, we generated volume regions of interest (ROIs) based on GM-voxels. 
Freesurfer provides parcellation of 34 cortical regions based on the Desikan atlas \cite{desikan} and 7 subcortical regions (Nucleus accumbens, Amygdala, Caudate, Hippocampus, Pallidum, Putamen, and Thalamus) 
\cite{fischl} for each hemisphere, thus leading to 82 ROIs in total. 
To quantify the connection strength from DTI, we performed eddy-current correction (FSL), and the Fiber Assignment by Continuous Tracking (FACT) algorithm \cite{mori} with 35 degrees angular threshold using Diffusion Toolkit along with TrackVis \cite{wang}. We then counted the number of streamlines between all pairs of defined regions of interest as connection strength using the UCLA Multimodal Connectivity Package (http://ccn.ucla.edu/wiki/index.php).
This led to an undirected weighted connectivity matrix $\mathbf{S}$ of size $N\times N$ with $N(N-1)/2$ 
connections per subject, where $N$ is 82. We also defined $\mathbf{L}$ as a $N\times N\times M$ matrix with the connectivity matrices $\mathbf{S}_k$ of $M=121$ subjects, where $k$ is the subject
index.

\subsection{Age prediction}
\label{sec:prediction}

Numerous machine-learning approaches could be applied for predicting features such as age. Instead, we focused on simple predictions that look at the correlation between the strength of each 
fiber tract, as measured by the number of streamlines and age. The correlation, ranging from -1 to 1, gives us a weight measure of how important each connection is for a feature 
leading to the group correlation matrix $\mathbf{C}$. The stronger a connection, given by the number of streamlines, the larger the impact on the predicted feature. Correlation matrix $\mathbf{C}$ is
defined as $C_{ij}=cov(L_{ij},A) / (\sigma_{L} \sigma_A) ~$, where $L_{ij}$ is the vector of connection strength between node $i$ and $j$ for all subjects, $A$ is the vector of ages ordered according to the $M$ subjects, and $cov$ and $\sigma$ stands for covariance and
variance respectively. 

In order to predict age, a predictor value $P$ is then given by the sum over all edges of the individual product between the correlation matrix $\mathbf{C}$ and the subject's matrix $\mathbf{S}$ as follows;
\begin{equation}
 P_k=\frac{1}{2} \sum_{i=1}^{N}\sum_{j=1}^N S_{ij}\times C_{ij}~.
\end{equation}
To compute predictor $P$, we divided our dataset in two groups; a training matrix $\mathbf{L}_R$ and a test matrix $\mathbf{L}_T$, where the former was used to compute 
$\mathbf{C}$ and the latter to compute $P$. Using the training data set, $P$ was then translated and re-scaled to map the age variable.
Additionally, we tested different ways of normalizing the group matrices $\mathbf{L}_R$ and $\mathbf{L}_T$; 1) No normalization, $\mathbf{L}$, 2) normalization between subjects, $\mathbf{L}'$, 
3) normalization within subjects, $\mathbf{L}''$. The last two are defined as
\begin{equation}
 L'_{ijk}=\frac{L_{ijk}}{\frac{1}{M} \sum_{i=j}^M L_{ijk}}~,
\end{equation}
\begin{equation}
 L''_{ijk}=\frac{L_{ijk}}{\frac{1}{2} \sum_{i=1}^N \sum_{j=1}^N L_{ijk}}~.
\end{equation}
Notice that in the previous equations $M$ is the number of subjects in each of the groups, and normalization is applied separately to $\mathbf{L}_R$ and $\mathbf{L}_T$.

\subsection{Simulations}

To test our age prediction, we used a bootstrapping approach. Testing $\mathbf{L}_T$ and training $\mathbf{L}_R$ matrices were created by randomly dividing the population matrix $\mathbf{L}$
at $50\%$ (half of the population for training or test), and then correlation matrix $\mathbf{C}$ and prediction vector $P$ for each division were computed. We used $100$ divisions to obtain extreme and average performances of our age prediction.

\section{Results}
\label{sec:results}

\begin{figure}[htb]

  \centering
  \centerline{\includegraphics[width=8.0cm]{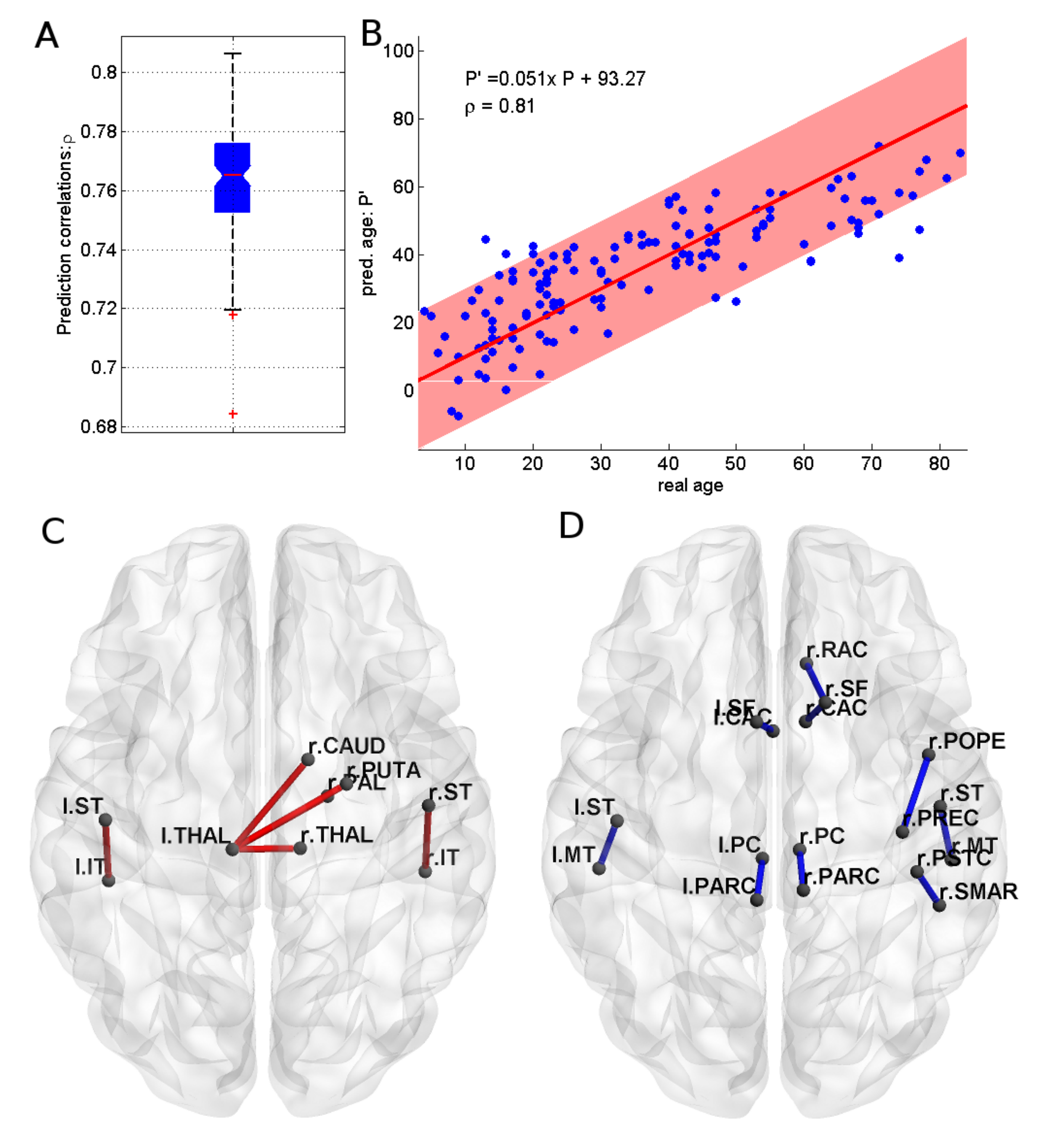}}
  \caption {Age prediction results. A) Correlation box plot of 100 prediction experiments by randomly dividing the population matrix in training and 
  test datasets. B) Best case from the 100 experiments where correlation between predicted and real ages is $\rho=0.81$, with $95\%$ confidence interval shown as shades.
  C) Highest $1\%$ of the edges with positive correlation with age. D) Lowest $1\%$ of the edges with negative
  correlation with age. Correlated edges were visualized with BrainNet Viewer (http://www.nitrc.org/projects/bnv/).}  \label{fig:results}

\end{figure}

Figure~\ref{fig:results} showed the results for age prediction experiments. For these, the entire database was divided in two matrices: training (60 subjects) and testing (61 subjects) matrices whose 
subjects were chosen randomly 100 times. For each of the 100 prediction iterations an age-edge correlation matrix $\mathbf{C}$ and predictor vector $P$ were computed. Figure~\ref{fig:results}A showed the
correlation coefficient distribution between the predicted ages and the real ages for the 100 experiments; the median was $\rho=0.774$ (see also Table~\ref{tab:correlation}) with $\rho=0.68$ and $\rho=0.81$ as
worst and best-case predictions, respectively. Figure~\ref{fig:results}B showed the best case prediction from the 100 experiments and the translation/re-scaling mapping equation of predictor vector $P$ for age.

Prediction results for the three normalization methods tested were shown in Table~\ref{tab:correlation}. The best performance was obtained without normalization, followed by normalization within subjects $L''$.

To infer which brain edges were most correlated with age prediction, the 100 estimated correlation matrices $\mathbf{C}$ were averaged and the most important edges with positive and negative 
correlations were extracted. Figure~\ref{fig:results}C showed the $1\%$ edges with the highest correlation with age. Similarly, Figure~\ref{fig:results}D showed the $1\%$ of the network edges with lowest negative 
correlation with age. These edges and their averaged correlations were listed in Table~\ref{tab:edges}.

\footnotesize
\begin{table}
\begin{center}
\footnotesize
\caption{Performance of age prediction. The best performance was obtained by non-normalization of the connectivity matrix. The table shows the first, second, and third quartiles from the correlation distribution of 100 prediction experiments for each normalization method.}  \label{tab:correlation}
\begin{tabular}{c c c c}
  \hline
    Normalization        &    $Q_1$   & $Q_2$     & $Q_3$   \\ \hline \hline
    $L$  &    0.7603    &  \textbf{0.774}     & 0.7878  \\
    $L'$  &    0.5604    &  0.599     & 0.6289  \\ 
    $L''$  &    0.6496    &  0.678     & 0.7095  \\ \hline
\end{tabular}
\end{center}
\end{table}
\normalsize


\section{Discussion}
\label{sec:discussion}
Our simple predictor estimated age of a subject given structural connectivity data with fair performance. The predicted age was highly correlated with the real age (best $\rho=0.81$). The top edges 
correlated with age included the previously reported areas: Regional efficiency, capturing regional information integration, was correlated with age positively in temporal and frontal regions 
and negatively in parietal and occipital regions previously \cite{gong}. Though it was hard to compare directly due to different atlases and different modalities, they also found superior and inferior temporal 
gyri, and superior frontal gyri as we did. The edges with negative correlations were partially matched with white matter integrity (Fractional Anisotropy, FA) changes over age in a voxel based DTI study \cite{stadlbauer2012}: 
notably, both superior frontal gyri, superior temporal gyri, anterior and posterior cingulate cortices. We note that to match ours with their results, more extensive investigation is required, because we only showed the top $1\%$ of edges.

Our method is quite simple but connected with machine learning concepts. The concept behind our method is the basic artificial neural networks, especially a radial basis function (RBF) network. 
Each RBF responded to a corresponding edge, and its learnable weight is a correlation coefficient we computed, capturing its influence on age. Statistically, this idea can be connected with the partial least 
square (PLS) regression \cite{wold}. PLS projects the original predictors (edges) into a lower dimensional space, using colinearity of each predictor variable with the feature variable (age) to be fitted. 
In our method, a simple Pearson correlation coefficient, that represents fitting quality of linear regression, captured colinearity. 
Then, a weighted sum of a subject's connection strength represented projection of the subject into the predicting score $P$, where the vector of correlation coefficients was projection mapping, while PLS combines them in a statistically meaningful way. 

\begin{table}
\begin{center}
\footnotesize
\caption{List of strongest top 1\% connections correlated with age}  \label{tab:edges}
\begin{tabular}{@{}l@{} c@{}}
  \hline
    Edge                                                       & $\bar{\rho}$   \\ \hline \hline
    Positive correlations                                      &  \\ \hline
    L. Thalamus -R. Thalamus.                                   & 0.3562  \\
    L. Superior temporal - L. Inferior temporal                 & 0.3534 \\
    ~~~~~gyrus.  &    \\         
    L. Thalamus - R. Caudate Nucleus.                           & 0.3135  \\
    L. Thalamus - R. Putamen.                                   & 0.3018  \\
    L. Thalamus - R. Pallidum.                                  & 0.2872  \\
    R. Superior temporal - R. Inferior temporal                 & 0.2902  \\
    ~~~~~gyrus.   &     \\     \hline
    Negative correlations                                      & $\bar{\rho}$ \\ \hline

    L. Posterior cingulate cortex - L. Paracentral              & -0.5905  \\
    ~~~~~gyrus.  &  \\
    L. Superior frontal - L. Caudal anterior           & -0.5613  \\
    ~~~~~cingulate cortex.  &      \\
    R. Superior temporal - R. Middle temporal gyrus.              & -0.5340  \\
    L. Superior temporal - L. Middle temporal gyrus.              & -0.5204  \\
    R. Posterior cingulate cortex - R. Paracentral              & -0.4797  \\
    ~~~~~gyrus.  & \\
    R. Superior frontal - R. Caudal anterior            & -0.4791  \\
    ~~~~~cingulate cortex. &  \\
    R. Parso percularis - R. Precentral gyrus.                    & -0.4518  \\
    R. Rostral anterior cingulate - R. Superior          & -0.4373  \\
    ~~~~~frontal gyrus.	& \\
    R. Postcentral gyrus - R. Supramarginal gyrus.                & -0.4324  \\
  \hline
\end{tabular}
\end{center}
\end{table}
We observed that non-normalization performed best in general, while between-subject normalization deteriorated its performance. First of all, we noted that this normalization was not standardization 
with the mean and standard deviation of each edge, but regularization of magnitudes to minimize effects of variability in edges' magnitudes. The vector of correlation coefficients captured the pattern of edges 
correlated with age. Because correlation coefficients captured the fitting quality (a clear relationship), not the magnitude of influence itself, we worried that a few edges whose overall magnitude is relatively larger than for other edges dominantly affected the prediction. So, we regularized edges with large average magnitudes by dividing the group mean of the edge (between-group normalization), and the individual sum of edges (within-group normalization). Though the between-group normalization does not affect correlation coefficients, it performed worse than the within-group normalization, which deteriorated correlation. Thus, we believed that not only the clear relationship captured by the correlation coefficients, but also the magnitude of the edge was important. In other words, edges with larger magnitudes (preponderantly larger variance) had greater roles in predicting brain maturity than edges with smaller magnitudes. 

Reducing the number of features is crucial in machine learning literature, including recursive feature elimination (RFE) \cite{mwangi2013}. Though we did not remove any edges, when its correlation coefficient 
is near zero, the edge can be considered as a removed feature. Using correlation coefficients for feature elimination was common \cite{dosenbach, yun2013}. 
Dosenbach et al. \cite{dosenbach} selected connections with 200 largest absolute values of correlation coefficients to predict age, which survived after Bonferroni correction. 
We note that such a selection with Bonferroni did not work for our dataset; we may employ a (less conservative) cluster-based correction method for correlation coefficients \cite{han}. 
A notably similar work with DTI and lifespan age prediction \cite{mwangi2013} used various white matter parameters including white matter integrity of voxels and support vector regression with RFE. Though the model showed slightly stronger correlation between the predicted age and the real age ($\rho \approx 0.89$), our method has the advantage 
of providing direct linkages between regions, which are useful to understand dynamic changes of brain connectivity during development. 

The method can be improved in various ways. First, the use of nonlinear fitting may help. A potential limitation is the use of a linear fit for converting the raw prediction measure $P$ to the age prediction, $P'$. While a linear fit yielded good results, exponential, logarithmic or other fits might be more suitable.  For example, we noticed an underestimation of age for subjects older than 50 years which links with earlier studies that found nonlinear trends during development for brain volume \cite{westlye,lebel2012}, white matter integrity \cite{westlye,lebel2012,mwangi2013}, and functional connectivity \cite{dosenbach}. 
Second, we may use partial correlation coefficients \cite{yun2013} to adjust group biases, or Spearman correlation coefficient to handle non-normality of connection strengths. Third, separating males and females may improve the performance as they may follow different developmental trajectories \cite{gong,sol}.

\section{Conclusion}
\label{sec:conclusion}
Using a simple predictor based on the influence of each fiber tract connection on a feature of interest, we were able to predict age of a subject given structural connectivity data with fair performance. 
This indicates that predicting other cognitive and general features (e.g. gender) might be derived from diffusion tensor imaging data of a subject. Also the proposed method is applicable to networks with other 
modalities including functional connectivity. We note that it can be improved in various ways still, including nonlinear mapping between a raw prediction score $P$ and a predicted 
age $P'$, and the use of partial correlation coefficients.

Besides the clinical and basic scientific applications, our results also raise an ethical issue. If anonymized DTI data of a subject and the metadata table of name, IQ, and age of all subjects is available, one 
might identify the name of the subject by finding table entries that match the predicted IQ and age. In other words, even without a key (identifier) that links entries of the MRI database with those of the clinical scores/metadata database, anonymized data could be linked to a specific person.

\noindent


\end{document}